# Characterization of timing jitter spectra in free-running mode-locked lasers with 340 dB dynamic range over 10 decades of Fourier frequency

Kwangyun Jung and Jungwon Kim*

*School of Mechanical, Aerospace and Systems Engineering, Korea Advanced Institute of Science and Technology (KAIST), Daejeon 305-701, Korea*
*Corresponding author: jungwon.kim@kaist.ac.kr*

We demonstrate a method that enables accurate timing jitter spectral density characterization of free-running mode-locked laser oscillators over more than 10-decade of Fourier frequency from mHz to tens MHz range. The method is based on analyzing both the input voltage noise to the slave laser and the output voltage noise from the balanced optical cross-correlator (BOC), when two mode-locked lasers are synchronized in repetition rate by the BOC. As a demonstration experiment, timing jitter spectrum of a free-running mode-locked Er-fiber laser with a dynamic range of >340 dB is measured over Fourier frequency ranging from 1 mHz to 38.5 MHz (Nyquist frequency). The demonstrated method can resolve different noise mechanisms that cause specific jitter characteristics in free-running mode-locked laser oscillators for a vast range of time scales from <100-ns to >1000-s.

Spectral purity of electronic and photonic oscillators is important for advancing various scientific and engineering applications such as high-precision synchronization [1], high-speed and high-resolution sampling and analog-to-digital converters [2], timing, time-keeping and navigation systems, clock distribution and communication networking equipment, signal measurement instrumentation, and radars and lidars, to name a few. For the optimization of oscillator performances, accurate characterization of phase noise and timing jitter of periodic signals generated from the oscillators is first required.

Recently it has been identified that femtosecond mode-locked lasers can serve as ultralow-noise photonic and electronic oscillators: mode-locked lasers can generate both optical pulse trains with extremely low timing jitter and, via proper optical-electronic (O-E) conversion process, microwave signals with extremely low phase noise. Recent measurements showed that free-running, passively mode-locked fiber lasers can generate both ~100-MHz repetition rate optical pulse trains and 10-GHz microwave signals with integrated timing jitter well below a femtosecond (when integrated from 10 kHz to >10 MHz Fourier frequency) [3-5].

Due to the intrinsically low level of timing jitter (e.g., lower than $10^{-4}$ fs$^2$/Hz at 10 kHz Fourier frequency [3,4]) and the equivalent phase noise (e.g., <-140 dBc/Hz at 10 kHz Fourier frequency for 10-GHz carrier [5]), the accurate measurement of timing jitter spectral density in mode-locked laser oscillators has been a challenge. Traditionally, the timing jitter of optical pulse trains was measured in an indirect way by characterizing the phase noise of microwave signals converted from the optical pulse trains via O-E conversion (typically, using fast (>GHz) photodetectors) [6]. After a bandpass filtering of one harmonic frequency component, the microwave signal is mixed in quadrature with low-noise electronic tracking oscillator, which can be performed by several commercially available signal source analyzers (such as [7]). Although this photodetection-based method is a simple method that can utilize microwave components and commercial instruments, its measurement dynamic range and resolution are insufficient for accurate characterization of mode-locked lasers. First, the measurement dynamic range and resolution are often limit by the shot noise and thermal noise in the photodetectors, which results in a typical measurement noise floor of ~ -140 to -150 dBc/Hz. In addition, the excess phase noise by amplitude-to-phase (AM-to-PM) conversion in the photodetection and microwave amplification can add unwanted timing jitter/phase noise, which is not part of real laser noise. This excess noise can be problematic for the accurate characterization of jitter in the low Fourier frequency when the laser is locked to a stable reference source.

As an alternative, the use of balanced optical cross-correlation (BOC) has recently enabled tens of attoseconds resolution characterization of the high-frequency timing jitter spectral density in mode-locked lasers up to the full Nyquist frequency [3,4,8]. The BOC-based timing detector method requires a low-bandwidth phase-locked loop (PLL), which uses two almost identical mode-locked lasers and locks the repetition rates of two lasers using a piezoelectric transducer (PZT)-mounted mirror driven by the output signal from the BOC via a proportional-integral (PI) servo controller. Due to the nonlinear optic nature of the BOC (e.g., using the sum-frequency generation for timing detection), the timing jitter spectral density of free-running mode-locked lasers can be measured down to the unprecedented $10^{-12}$ fs²/Hz (equivalently, <-220 dBc/Hz phase noise at 1-GHz carrier) level *outside* the locking bandwidth when monitoring the BOC output. In order to analyze the jitter spectra of free-running oscillators as broad range as possible, the laser locking bandwidth is usually kept as low as possible. The typically achievable lowest locking bandwidth is ~1 kHz range (as shown in [3,4,8]), which sets the lower bound limit of Fourier frequency in jitter characterization of free-running mode-locked lasers. Thus, the timing jitter spectra of free-running mode-locked lasers in the low (e.g., <1 kHz) Fourier frequency could not be characterized accurately so far.

In this Letter, we identify that the BOC jitter measurement range can be greatly extended to the much lower Fourier frequency (e.g., mHz and below) by analyzing the PZT driving voltage signal, which carries the frequency noise information between the two lasers. Our demonstration measurement shows that the jitter spectrum from 1 mHz to 38.5 MHz Fourier frequency is possible with a jitter spectrum range from ~$10^{-10}$ fs²/Hz to ~$10^{25}$ fs²/Hz (equivalent single sideband phase noise of -205 dBc/Hz to +140 dBc/Hz at 1-GHz carrier), which shows that jitter spectral density of free-running oscillators spanning >340 dB range can be accurately characterized over more than 10-decade of Fourier frequency. This method can completely resolve different noise mechanisms that cause specific jitter spectra slopes and shapes in free-running mode-locked laser oscillators from <mHz to the full Nyquist frequency ranges (which is, from typical <100-ns to >1000-s in time scale).

Figure 1 shows the schematic of the proposed jitter spectrum measurement method. Two almost identical, but independent, mode-locked lasers are used as a reference oscillator (Laser 1 in Fig. 1) and a slave oscillator with a repetition-rate tuning mechanism such as a PZT-mounted mirror (Laser 2 in Fig. 1). In addition to the traditional way of monitoring the error signal from the BOC output, which directly shows the timing jitter (and equivalent phase noise) of free-running oscillators *outside* the locking bandwidth, we also monitor the input voltage signal to the PZT in the slave oscillator (Laser 2 in Fig. 1), which effectively acts as a voltage-controlled oscillator (VCO) in the PLL [9]. Just like the electronic VCO in the typical PLLs, the repetition rate of the slave oscillator (Laser 2) is tuned as a function of the input voltage to the PZT, $f_R = f_{R0} + KV$, where $f_R$ is the resulting repetition rate, $f_{R0}$ is the center repetition rate without the PZT input signal, $K$ is the gain of the VCO (Hz/V), and $V$ is the voltage applied to the actuator (PZT). Thus, when the repetition rates of two lasers are locked, the input voltage to the PZT carries the frequency noise information between the two lasers *inside* the locking bandwidth. By using the definition between the phase noise spectral density and the frequency noise spectral density, $S_{\Delta\varphi^2}(f) = S_{\Delta f^2}(f)/f^2$, where $S_{\Delta\varphi^2}(f)$ is the phase noise spectral density (in rad²/Hz) at the fundamental repetition rate $f_R$, $S_{\Delta f^2}(f)$ is the frequency noise spectral density (in Hz²/Hz) and $f$ is the Fourier frequency (which is, the offset frequency from the carrier), one can convert the measured voltage noise power spectral density of the PZT input signals to the equivalent phase noise power spectral density at the carrier frequency of fundamental repetition rate ($f_R$). Finally, the timing jitter power spectral density $S_{\Delta t^2}(f)$ (in s²/Hz) can be obtained by $S_{\Delta t^2}(f) = S_{\Delta\varphi^2}(f)/(2\pi f_R)^2$. As the measurement result shows the power spectral density of relative jitter between the two almost identical, independent free-running lasers, the contribution from one laser is obtained by dividing the measured power spectral density by two. Therefore, by combining the BOC output noise spectrum *outside* the locking bandwidth and the PZT input voltage noise spectrum *inside* the locking bandwidth, as shown in the insets of Fig. 1, one can fully reconstruct the timing jitter spectra of free-running mode-locked laser oscillators over a wide range of Fourier frequency.

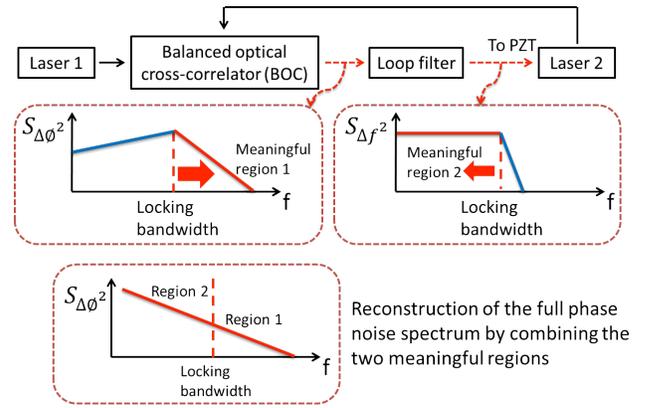

Fig. 1. Timing jitter measurement principles. Laser 1 (reference oscillator) and Laser 2 (slave oscillator) are almost identical, but independent, mode-locked lasers.

Figure 2 shows the schematic of demonstration experiment. For mode-locked lasers under test, two almost identical, independently operating, nonlinear polarization rotation (NPR)-based, 77-MHz repetition rate, passively mode-locked Er-fiber lasers are used. They operate in stretched-pulse regime at +0.002 ps² intra-cavity dispersion. They employ sigma-cavity free-space sections with PZT-mounted mirrors for the repetition-rate locking. The maximum optical path length compensation by the used PZT is 4.4-µm by the round-trip beam path. It corresponds to 87-Hz tuning range for the used 77-MHz repetition rate. Note that, although a PZT is used in this experiment, other methods for the repetition rate tuning, such as PZT fiber stretchers, motorized stages, electro-optic modulators [10] or pump modulation [11], can be employed as well. The outputs from the two lasers are combined by a polarization beam splitter (PBS in Fig. 2) and applied to the BOC. The BOC used in this work uses the same configuration as shown in [3] with a type-II phase-matched periodically poled KTiOPO$_4$ (PPKTP) for the second-harmonic generation (SHG) and for the fixed group delay between the two polarization states. The output from the BOC is fed back to the PZT via a PI servo and a high-voltage PZT amplifier, and used for the repetition rate locking between the two lasers. When locked, the voltage noise spectra of the BOC output and the PZT input are characterized. As the PZT input voltage is very high (~100 V), we measure voltage signal before the PZT amplifier. The PZT input voltage can be calculated from it, because the amplifier gain is constant inside the locking bandwidth. An RF spectrum analyzer (Agilent, E4411B) and an FFT spectrum analyzer (Stanford Research Systems, SR770) are used for measuring the spectral densities in 81 kHz – 38.5 MHz (Nyquist frequency) and 0.76 Hz – 81 kHz ranges, respectively. For the Fourier frequency below 0.76 Hz, we used a data acquisition board (National Instruments, PCI-6221) and performed an FFT using MATLAB to obtain the power spectral densities. Due to the temperature drift in the laboratory, the repetition rate locking could be maintained for ~25 minutes. As a result, the measurement range was limited to ~1 mHz in Fourier frequency. Provided a longer-range actuator in the laser cavity, the measurement range could be extended to a much lower Fourier frequency.

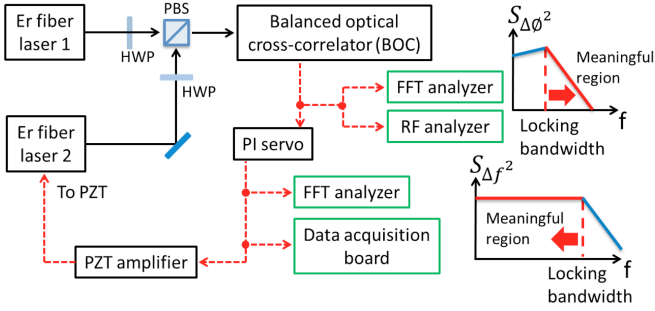

Fig. 2. Experimental setup. PBS, polarization beam splitter; HWP, half-wave plate; PI, proportional-integral; $S_{\Delta\varphi^2}$, phase noise spectral density; $S_{\Delta f^2}$, frequency noise spectral density.

Once the voltage noise spectra at the BOC output and the PZT input are obtained, they should be converted to the equivalent timing jitter spectra. For the BOC part, we detune the repetition rates of two lasers and measure the slope in fs/V unit. The measured slope is 0.43 fs/mV. By multiplying the voltage noise spectrum at the BOC output with the measured slope, timing jitter spectral density of free-running lasers (in fs²/Hz) is directly obtained outside the locking bandwidth. Next, for calibrating the measured PZT input voltage, the laser repetition rate tuning sensitivity (in Hz/V) should be measured. We modulate the PZT with known voltage amplitude and measure the repetition rate change. The measured slope is 1.1 Hz/V. Using the measured slope, frequency noise spectral density at the fundamental repetition rate is obtained, which can be converted to the equivalent timing jitter spectral density inside the locking bandwidth. We also measured the linearity of the used PZT, and it showed up to ~3 % nonlinearity in the voltage range used for jitter measurement over 25 minutes.

Figure 3 shows the measured timing jitter spectral density of the free-running mode-locked Er-fiber laser under test. Curves (i) and (ii) are the measurement results from the PZT input and the BOC output, respectively. The used PLL locking bandwidth is ~12 kHz. The PZT measurement [curve (i) in Fig. 3(a)] shows slight overshoot in the 3 kHz – 10 kHz range by the feedback loop. Note that, in principle, this distortion of free-running jitter spectrum near locking bandwidth can be corrected by intentionally using two sufficiently different locking bandwidth conditions and combining the two results. However, experimentally we found that the usable locking bandwidth with BOC method is in a rather limited range, and could not completely remove the slight bump caused by the PLL. We used the worst case measurement result for Fig. 3 to show the upper limit of the jitter spectrum. Since the used locking bandwidth is ~12 kHz, the PZT input noise [curve (i) in Fig. 3(a)] and the BOC output noise [curve (ii) in Fig. 3(a)] represent the jitter spectral density of the free-running laser below and above 12 kHz Fourier frequency, respectively. From ~20 kHz to ~8 MHz Fourier frequency, the measured jitter spectrum follows the amplified spontaneous emission (ASE) quantum noise-limited random walk nature (-20 dB/decade slope) of mode-locked lasers [12]. At higher Fourier frequency above 8 MHz, the measurement is limited by the BOC noise floor of ~2×10⁻¹⁰ fs²/Hz (i.e., equivalent phase noise of -205 dBc/Hz at 1 GHz carrier). In the lower Fourier frequency below 1 kHz [curve (i) in Fig. 3(a)], strong noise spikes are observed from 30 Hz to 1 kHz Fourier frequency range, which are caused by the AC power (60-Hz and harmonics) and acoustic noise coupled to the lasers. One of the most notable findings in this measurement is a steep slope (~ -60 dB/decade) observed in the 0.1–1 Hz Fourier frequency range. Although the exact origin of this steep slope in jitter spectrum is unclear yet, our initial analysis suggests that the atmospheric fluctuations [13] in the ~24-cm long free-space section of lasers can contribute to this level of jitter spectrum.

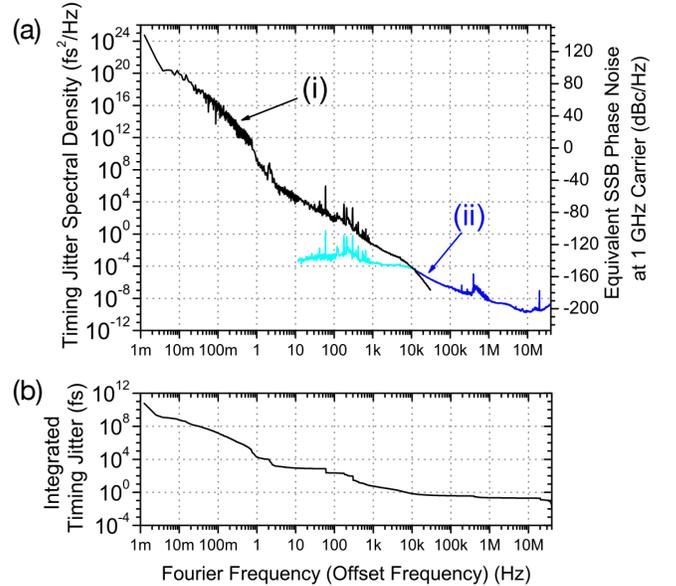

Fig. 3. (a) Timing jitter spectrum measurements. Curve (i) and (ii) are obtained from the PZT input noise and the BOC output noise, respectively. (b) Integrated rms timing jitter.

Figure 3(b) shows the rms timing jitter progressively integrated from 38.5 MHz down to 1 mHz Fourier frequency (i.e., from 26 ns to 1000 s integration time). At 10 kHz (0.1 ms integration time), the integrated timing jitter is 0.71 fs (rms) mostly originated from the ASE quantum noise. After strong acoustic noise-induced spikes in the 100 Hz – 1 kHz and the rapid divergence of timing/phase around 2 Hz, the integrated jitter reaches 20 ps (rms) at 1 Hz (1-s integration time). Note that the projected integrated jitter is only 71 fs when assuming that the jitter is limited only by the ASE quantum noise with -20 dB/decade slope. However, due to much steeper increase in jitter slope and technical noise sources such as 60-Hz AC power and acoustic noise coupling in the low Fourier frequency range (1 Hz – 1 kHz), the actual measured jitter is ~280 times higher in the 1-s time scale. The final timing jitter reaches 0.1 ms level at 1 mHz Fourier frequency (~17 minutes integration time). Figure 4 shows the fractional frequency instability in terms of Allan deviation calculated from the measured timing jitter spectrum in Fig. 3(a) [14]. The minimum frequency instability reaches $8.8 \times 10^{-12}$ at 1.3 ms averaging time. As the averaging time increases, due to the acoustic noise-coupled peaks and rapid divergence in jitter spectrum, the frequency instability increases to $3.4 \times 10^{-11}$ and eventually to $1.1 \times 10^{-8}$ at 1 s and 308 s averaging time, respectively.

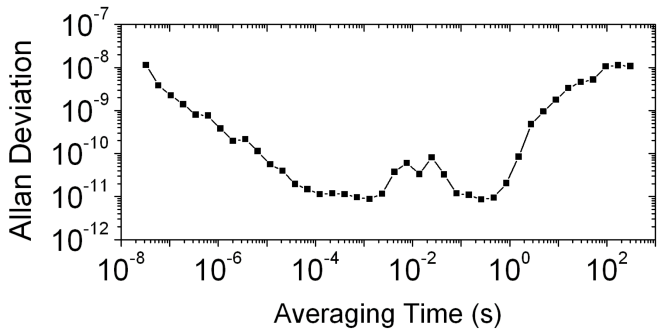

Fig. 4. Allan deviation of the free-running mode-locked Er-fiber laser calculated from the timing jitter spectral density data in Fig. 3(a).

We also tested the validity of our measurement method by comparing it with the direct photodetection-based phase noise measurement method. We built a direct photodetection-based setup as shown in [3] and mixed the 1-GHz components (13th harmonic) in quadrature using a microwave mixer. The mixer output and the PZT input signals are monitored in a similar way proposed in this work. The results are shown in Fig. 5. In the high Fourier frequency (>10 kHz), the BOC-based measurement [curve (a)] has >60 dB higher measurement dynamic range than the mixer-based measurement [curve (b)]. Below 4 kHz Fourier frequency, the two measurements are almost identical. For free-running oscillators, the integrated jitter rapidly increases in the low Fourier frequency and the added jitter in the O-E conversion is much smaller than the laser jitter. Note that the noise floor of BOC method is -170 dBc/Hz even at 1 Hz, which indicates that the BOC method can characterize stabilized laser sources as well.

In summary, we have demonstrated a measurement method of timing jitter spectral density in free-running mode-locked lasers with greatly extended Fourier frequency range down to the mHz level by analyzing the PZT driving voltage signal. The demonstrated method paves a new way for directly observing the timing jitter and drift in free-running oscillators with attosecond resolution over the entire Nyquist frequency, which could not be thoroughly investigated so far. Prior knowledge of complete jitter spectrum in free-running laser oscillators can also greatly improve the design of optimal servo controller (loop filter) transfer function for achieving the lowest possible residual jitter in the laser-laser or laser-microwave synchronization.

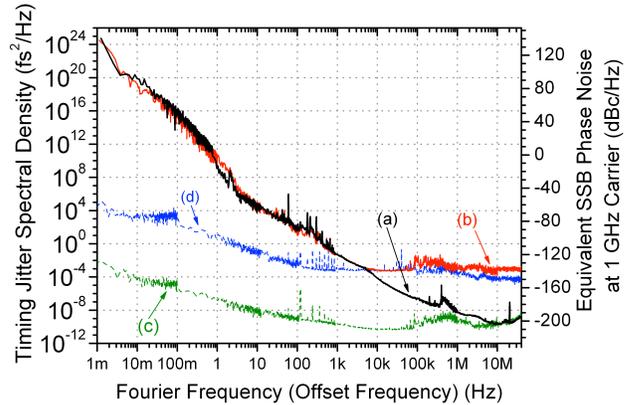

Fig. 5. Comparison of (a) BOC-based measurement result and (b) photodetection-based measurement result. (c) BOC method background noise (0.17 fs integrated jitter [1 mHz-38.5 MHz]). (d) Photodetection method background noise (68 fs integrated jitter [1 mHz-38.5 MHz]).

This research was supported by the National Research Foundation (NRF) of South Korea (under grant 2012R1A2A2A01005544).